\documentclass[preprint]{elsarticle}

\usepackage{graphicx}
\usepackage{dcolumn}
\usepackage{bm}

\journal{Chemical Physics}

\begin{document}

\begin{frontmatter}

\title{%
Collisional state-changing of OH$^-$ rotations  by interaction with Rb atoms in cold traps%
\tnoteref{mytitlenote}}
\tnotetext[mytitlenote]{%
This work is affectionately dedicated to Prof. Andrey Belyaev, on the occasion of his 60$^{th}$ birthday:
to an excellent scientist and a very dear friend}

	\author[usal]{L. Gonz{\'a}lez-S{\'a}nchez}
	\author[innsbruck]{F. Carelli}
        \author[innsbruck,pisa]{F.A. Gianturco\corref{mycorrespondingauthor}}
\cortext[mycorrespondingauthor]{Corresponding author: F.A. Gianturco. Email: francesco.gianturco@uibk.ac.at}
	\author[innsbruck]{R. Wester}

\address[usal]{Departamento de Qu{\'i}mica F{\'i}sica, Universidad de Salamanca, Pza. de los Ca{\'i}dos sn, 37008,  Salamanca, Spain}
\address[innsbruck]{Institut f\"ur Ionenphysik und Angewandte Physik, The University of Innsbruck, Technikerstr. 25, A-6020, 
	Innsbruck, Austria}
\address[pisa]{Scuola Normale Superiore, Pzza de Cavalieri 7, I-56126, Pisa, Italy}

\begin{abstract}
We employ an accurate, {\it ab initio}  potential energy surface (PES) which describes the electronic
 interaction energy between the molecular anion OH$^-$ ($^1\Sigma^+$) and the neutral rubidium atom Rb ($^2S$),
to evaluate the elastic and inelastic cross sections over a range of energies representative of the conditions of low-T
experiments in MOT traps, when combined with laser-cooled rubidium gas. The system is considered to be in its vibrational ground
state, while the first four rotational levels are taken to be involved in the cooling and heating collisional processes
that are computed here. 

The corresponding cooling and heating rates up to about 35 K are obtained from the calculations and compared with the recent results in a similar experiments, where He was the partner atom of the current anion.
\end{abstract}


\end{frontmatter}

\section{Introduction}
Experimental and computational advances in the study of the dynamics of cold molecules have allowed the opening up of several new avenues in 
various branches of the physical and chemical sciences \cite{Carr-etal:09, Duliew-Gabbanini:09,Schnell-Meijer:09}.
This variety of novel developments has shown up in different areas that range from fundamental precision data
\cite{Loh-etal:13}, quantum information processing \cite{Andre-etal:06}, quantum control of chemical reactions
\cite{Quemener-Julienne:12}, even as far as yielding  a more realistic analysis of the cold molecular interstellar medium \cite{Gerlich-etal:12}.

One of the important  consequences for this new investigative directions has been the development of new research fields
which deal exclusively with the formation, controlled preparation and storage of cold molecules for which the 
translational, rotational and vibrational or hyperfine energies can correspond to temperatures either below 1 K or already
well below the conventional room temperature values of $\sim$ 300 K.

In the case of molecular ions it has been found, already for a while, that collisional cooling is a very good general method
to cool such ions, whereby translational temperatures down to a few millikelvins can be reached by sympathetic cooling of molecular 
ions with laser-cooled atomic ions trapped jointly in a radiofrequency (rf) trap \cite{Schiller-Roth:09}.
However, because of the nature of the long-range interaction between ions, in the case of 
molecular partners such experimental setups do not lead to 
internal quantum state cooling . A possible alternative is therefore provided by cooling molecular ions 
in collisions with cold neutral atoms, where the occurrence of much closer encounters between 
the ions in the trap and the
cloud of laser-cooled atoms can in principle allow for a more efficient population reduction( i.e.cooling)
 of rotational and vibrational states through 
the increased frequency of collisional energy transfers over short time intervals. When using standard cryostats employing He,
the experiments are limited to temperatures above about 4 K as the coldest conditions reached in the traps
\cite{Pearson-etal:95}.

In order to reach  translational temperatures in the mK regime, and also cool the internal degrees of freedom to a few K or
below, experiments have striven to use hybrid atom ion traps (HAITrap) for laser cooled neutral atoms and molecular ions
\cite{Hudson:09}. Such arrangements consist of an rf-trap superimposed with a magneto-optical trap (MOT), so that the
ions are immersed in a cloud of laser-cooled heavier atoms. Using such an arrangement, therefore, added cold atoms have been
used to internally cool translationally cold molecular ions \cite{Rellergert-etal:13}. Hence, one might expect that, in order
to investigate the effects of sympathetic cooling and to further study chemical reactions in cold and ultracold ($<$ 1 millikelvin)
temperature regimes, the hybrid set-ups in traps described above should provide an interesting avenue of investigation.

Because of the crucially needed presence of efficient collisional conditions in such experiments, it becomes indeed vital to be able to
link the experimental  findings with realistic estimates of the collision loss channels and the relative efficiency of the primary sympathetic 
cooling approach, the latter being controlled by the size of the cross sections for the elastic (purely translational) collisional channels, with respect to the size and temperature behaviour of the state-changing scattering processes.

Recent experiments have been already discussing  a specific molecular anion (OH$^-$) that was made to interact with
a laser-cooled cloud of rubidium atoms \cite{Deiglmayr-etal:12}. The results were dedicated to the analysis of the hybrid
setup and the various ion trap conditions: they were able to estimate the relative temperature of the OH$^-$ + Rb
system in the experiment to be 400$\pm$200 K. They further estimated the value of the inelastic rate coefficient to be
2$^{+2}_{-1}\times$ 10$^{-10}$ cm$^{-3}$/sec, which therefore turns out to be significantly smaller than the one given by 
the Langevin model estimate: 4.3$\times$10$^{-9}$ cm$^{-3}$/sec\cite{Deiglmayr-etal:12}.

Throughout that work it was clearly indicated that knowledge of the collisional rates associated with both the translational
cooling of OH$^-$ and with the inelastic processes leading to its internal rotational cooling by collisions with the laser-cooled
rubidium atoms of the cloud would be an important ingredient for estimating the loss rates and the final production of internally cooled, 
and possibly also state-controlled OH$^-$ molecules. To that end, therefore, we are embarking here in a detailed calculation of the
rotationally inelastic cross sections, followed by the evaluation of the corresponding rates , for  OH$^-$ internal states
of $j$=1, 2 and 3 down to all the accessible channels. At the same time, we are evaluating the excitation cross sections
 from the j$_{initial}$=0, 1 and 2, as well as the total elastic cross sections , the latter being linked to the collisional efficiency of
the translationally cooling channels within an experimental set up of  Rb atoms  uploaded into the ion trap \cite{Deiglmayr-etal:12}.

In a very recent study of an experimental and computational analysis of rotational state-changing cold collisions for the
same target anion OH$^-$ but interacting with a buffer gas of He atom \cite {Hauser-etal:15}, it was found that
the internal and translational temperatures were brought down to  around 10-20 K. In the present study we are therefore 
carrying out calculations over the same range of temperatures to emphasize as much as possible a close comparison between 
the collisional energy-transfer rates of a buffer gas like He and those which are expected  by Rb collisional cooling experiments.

The following section summarizes the main features of the previously computed potential energy surface (PES) 
\cite{Gonzalez-Sanchez-etal:08} and also provides an outline of the quantum coupled-channel method employed to generate the relevant
cross sections \cite{Lopez-Duran-etal:08}.The next section 3 reports in detail our present results and compares them with both earlier experiments and similar systems.Our final conclusions are given by section 4.

\section{The computational machinery}
\subsection{The ab initio interaction forces}

The ground-state electronic structure of the RbOH$^-$ system is given by the interaction between the OH$^-$ electronic
ground state with term symbol $^1\Sigma^+$ and the neutral rubidium atom, the ground state of which is represented by
$^2S_{1/2}$. It thus follows that the linear configuration of the complex has a term symbol of $^2\Sigma^+$. Since
we are considering the target molecule to be a rigid rotor in its ground vibrational state, the spatial features of that 
electronic interaction have been described by two Jacobi coordinates only (R,$\Theta$,r$_{eq}$). Here r$_{eq}$, indicating the equilibrium geometry of the anionic partner, was 
taken to be 0.79 \AA \cite{Gonzalez-Sanchez-etal:08} and the variables (R,$\Theta$) were used on a discrete grid within the
radial interval of [1.9-18.00] \AA, with a variable length of the radial step (see ref. \cite{Gonzalez-Sanchez-etal:08}
for further details). The angular variable ranged, as expected, over the [0$^\circ$-180$^\circ$] interval with $\Delta\Theta$=15$^\circ$.

The {\it ab initio} calculations involved an all-electron correlation treatment with the exception of the ($^1S^2$) 
oxygen core. All the computational details are given in \cite{Gonzalez-Sanchez-etal:08}, while we present
here simply an overall view of those results to better explain and discuss  the ensuing dynamics of the quantum collisions.
Just as a further reminder, in the calculations we followed the full counterpoise procedure \cite{Boys-Bernardi:70}
to account for the effects of basis set superposition errors (BSSE).

 \begin{figure*}
 \includegraphics[scale=0.7]{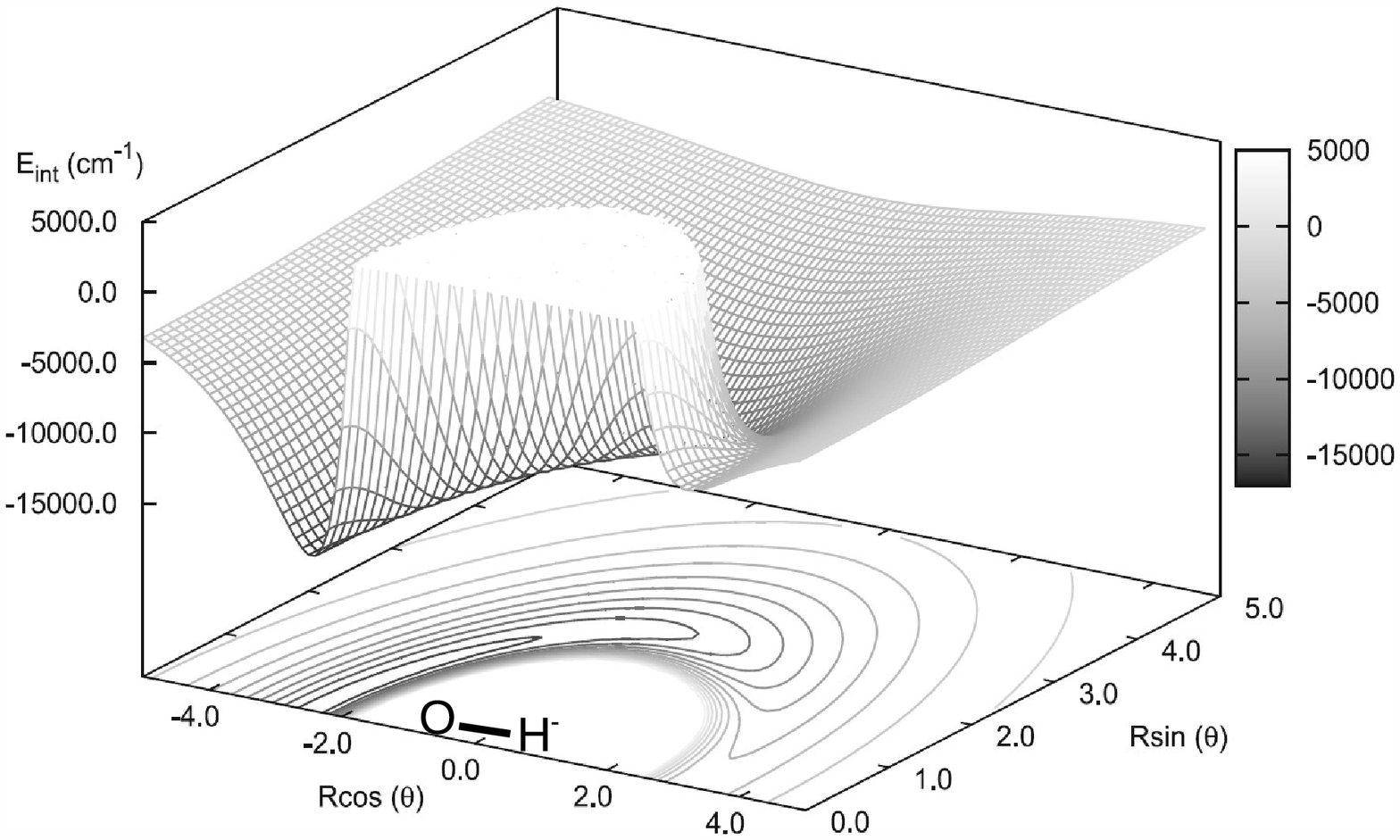}%
 \caption{\label{fig:pes}Computed interaction potential between OH$^-$($^1\Sigma^+$) and Rb($^2$S) atoms. The plot
 presents a 3D view of the energy surface as a function of the two Jacobi coordinates (R,$\Theta$). See main text
 for further details (adapted from \cite{Gonzalez-Sanchez-etal:08}).}%
 \end{figure*}

The 3D plot presented by Figure \ref{fig:pes} describes pictorially the shape of the full PES of the title system
and provides us with useful information on the overall features of the present interaction potential. One notices clearly, 
in fact, that the (OH$^-$-Rb) complex gives rise to a strongly bound system. The global energy minimum is located on the
side of the oxygen atom, for $\Theta$=180$^\circ$ in our representation, at a distance from the center-of-mass (c-o-m) of
2.44 \AA: it carries a minimum energy value of 1.65$\times$10$^4$ cm$^{-1}$ which is clear indication of the presence 
of a strong chemical bond. The overall PES is also strongly anisotropic (i.e. $\Theta$-dependent) and exhibits a second, 
less strong minimum for $\Theta$=0$^\circ$ that is still well marked but smaller than that of the other linear configuration:
6.27$\times$10$^3$ cm$^{-1}$. Since the OH$^-$ carries a negative charge and a permanent polarizable moment, and the Rb
atom can be viewed as a polarizable dielectric we further added on the following long-range expansion:

\begin{eqnarray}
	V_{LR}(R,\Theta) = C_{4,0} \frac{f_d(R)}{R^4} + C_{5,1} \frac{f_d(R)}{R^5}\cos\Theta
	\label{eq:one}
\end{eqnarray}

In eq.\ref{eq:one} $f_d(R)$ describes radial damping functions which are used to extend the 2D-fitting to the external
radial regions, as described in ref \cite{Gonzalez-Sanchez-Gianturco:07}. The additional dipole-polarizability term is included via
 the following expression:

\begin{eqnarray}
	C_{5,1}\sim 2 \alpha_0 \mu
	\label{eq:two}
\end{eqnarray}

 \begin{figure}
	 \includegraphics[scale=0.7]{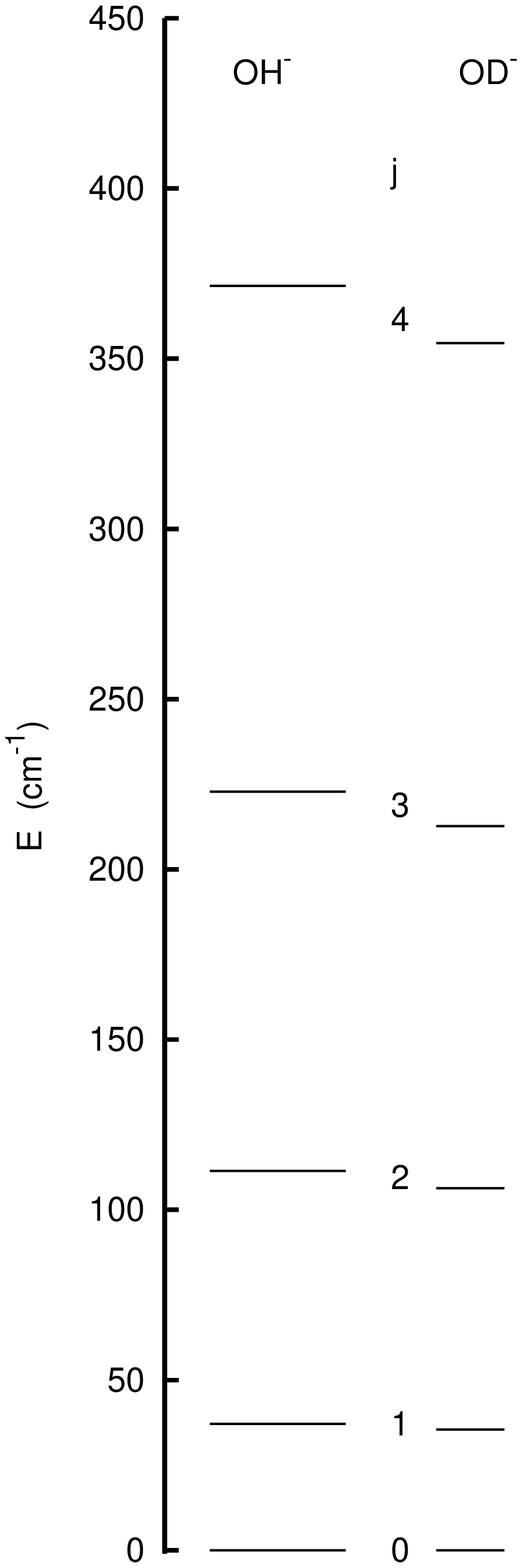}%
 \caption{\label{fig:states}Energy level diagrams for the lowest rotational levels of the OH$^-$ and of its isotopic variant.}%
 \end{figure}

 The lowest rotational states of the OH$^-$ target ion are represented in Fig. \ref{fig:states}. In the following we shall be considering transitions involving the first four rotational states of the target anionic molecule.

\subsection{The multichannel scattering equations}

The usual time-independent formulation of the close coupling (CC)
approach to quantum inelastic scattering  will not be  repeated
here in detail (see for example Ref. \cite{taylor} for a general formulation
and Ref. \cite{Arthurs-Dalgarno:60} for the discussion of a specific case). Only a short
outline is given below.

The time-independent scattering state of a system made of weakly interacting
partners can be expanded in terms of diabatic (asymptotic) target eigenstates:
\begin{equation}
\Psi ^{i,+}(R,\underline{x})=\sum _{f}\, F_{i\rightarrow
f}(R)*X_{f}(\underline{x})\label{channel_expansion}
\end{equation}
where $i$ labels the (collective) initial state of the colliding
partners and the $X_{f}$ are the eigenstates of the isolated molecule
(channel eigenstates). The $F_{i\rightarrow f}$ are the channel components
of the scattering wavefunction which have to be determined by solving
the Schroedinger equation subject to the usual boundary conditions
\begin{equation}
F_{i\rightarrow f}(R)\rightarrow \delta _{if}h^{(-)}(R)-S_{fi}h^{(+)}(R)\, \, \,
\, \textrm{as }\, \, R\rightarrow \infty \label{boundary}
\end{equation}
where $f$ denotes a channel which is asymptotically accessible at
the selected energy (open channel) and $h^{(\pm )}$ is a pair of
linearly independent free-particle solutions.  
Usually, numerically converged scattering observables are obtained by retaining
only a finite number of discrete channels in eq.(\ref{channel_expansion}). One
thus gets a set of $M$ coupled differential equations for the $F_{i\rightarrow
f}$ unknowns (which form a matrix solution
$\boldsymbol \Psi $)  subject to the regularity conditions of each solution
at the origin ($F_{i\rightarrow f}(0)=0$) and to the boundary conditions given
by eq.(\ref{boundary}). 

In the case where no chemical modifications are induced into the
target by the impinging  atom (as is in our case), the total scattering wavefunction can be
expanded in terms of asymptotic target rotational 
eigenfunctions (within the rigid rotor approximation) which are taken to be
spherical armonics and whose eigenvalues are given by
$Bj(j+1)$. Here the B value was taken to be 18.5701 cm-1,as given in our earlier work.
Following Ref. \cite{Arthurs-Dalgarno:60} the channel components of eq
(\ref{channel_expansion}) are therefore expanded into products of
total angular momentum eigenfunctions
and of radial functions.
These radial functions are in turn the elements of the solutions matrix which
appear in the  familiar set of coupled, second order homogeneous differential
equations:
\begin{equation}
\left\{
\frac{d^{2}}{dR^{2}}+\mathbf{k}^{2}-\mathbf{V}-\frac{\mathbf{l}^{2}}{R^{2}}
\right\} \boldsymbol \Psi=0\label{Close-Coupling}
\end{equation}
where $[\mathbf{k}^{2}]_{ij}=\delta _{ij}2\mu (E-\epsilon _{i})$
is the diagonal matrix of the asymptotic (squared) wavevectors and
$[\mathbf{l}^{2}]_{ij}=\delta
_{ij}l_{i}(l_{i}+1)$
is the matrix representation of the square of the orbital angular
momentum operator. 
This matrix is block-diagonal with two sub-blocks that
contain only even values of $(l'+j')$ or only odd values of
$(l'+j')$. 

The scattering observables are obtained in
the asymptotic region where the LogDerivative matrix has a known form in terms
of free-particle solutions and unknown mixing coefficients. For example, in the
asymptotic region the solution matrix can be written in the form
\begin{equation}
\Psi (R)=\mathbf{J}(R)-\mathbf{N}(R)\, \mathbf{K}
\end{equation}
where $\mathbf{J}(R)$ and $\mathbf{N}(R)$ are matrices of Riccati-Bessel
and Riccati-Neumann functions.
Therefore, at the end of the propagation one uses the LogDerivative matrix to
obtain the $\mathbf{K}$ matrix by solving the following linear system
\begin{equation}
(\mathbf{N}'-\mathbf{Y}\, \mathbf{N})\, \mathbf{K}=\mathbf{J}'-\mathbf{Y}\,
\mathbf{J}\label{YtoK}
\end{equation}
and from the $\mathbf{K}$ matrix one gets easily the S-matrix and the cross
sections.  
We have recently published an algorithm that modifies
the variable phase approach to solve that problem, specifically
addressing the latter point\cite{our-cpc} and we defer the interested reader to
that reference for further details.

In the present calculations, therefore, we have used multipolar potential terms up to $\lambda_{max}$= 9 in eq. \ref{eq:ten}
\begin{eqnarray}
	V(R,\Theta) = \sum_\lambda V_\lambda (R) P_\lambda (\cos\theta)
	\label{eq:ten}
\end{eqnarray}
Here the $\lambda$ $V_\lambda (R)$ describe the relative strength of the anisotropy for each multipolar term. We have extended 
the radial range of integration out to 15,000  \AA  using a total of 30,000 integration steps. The total number of
total angular momentum values was extended up to $J_{max}$= 250, while the number of rotational channels always included
at least 15 closed channels above the last open channel at the given collision energy . The accuracy of the above fitting of the 
ab initio raw points ranged from a few  inverse cm.in the well regions and the long-range parts to about 10 inverse cm. in the repulsive walls regions.

\section{Results and discussion}

In order to provide some preliminary results which can give us a better understanding of the collision-induced state-changing
events in an hybrid arrangement with negative ions and laser-cooled neutral atoms, we have carried out the calculations 
involving the first (lowest) four rotational levels of OH$^-$ and included translational energies that would allow us
to generate the corresponding rates up to about 35 K. These values are substantially lower than the translational temperatures
being achieved thus far with Rb atoms in either past  experiments \cite{Hauser-etal:15} or in the ongoing experiment of the
Heidelberg-Innsbruck collaboration \cite{Wester-Weidemueller}. However, since such temperatures are hopefully those aimed at by 
the current efforts \cite{Wester-Weidemueller}, we thought it important to accurately assess in our numerical experiment the final values expected
for the optimal experimental setup. Approximate extrapolations to the  higher temperatures that are being achieved at the intermediate stages of the experiments will  be presented and 
discussed at a later stage, together with a detailed analysis of the dynamical effects of isotopic changes in the molecular partner.

\subsection{Partial inelastic cross sections}
\begin{figure}
\includegraphics[scale=0.7]{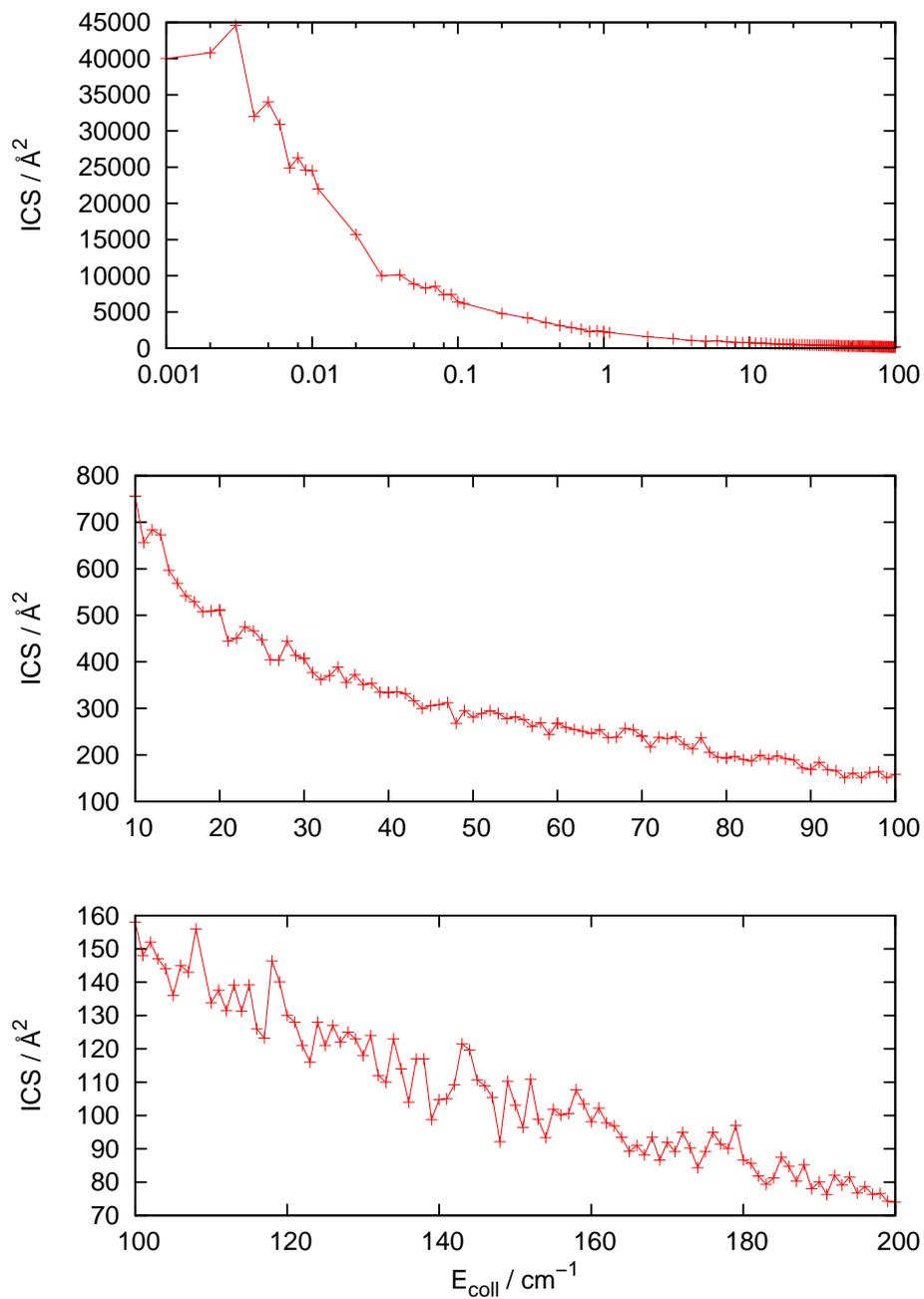}%
\caption{\label{fig:cooling1to0}Computed integral inelastic cross sections over different energy ranges for the (j=1$\rightarrow$j=0)
state changing process. See main text for further details.}%
\end{figure}

We report in figure \ref{fig:cooling1to0} the energy dependence of the state-changing cross sections for OH$^-$ in collision with
rubidium atoms. The three panels describe three different energy ranges: the top panel covers the dependence of the 
(j=1$\rightarrow$j=0) partial cross sections from about 1 millicm$^{-1}$ to 100 cm$^{-1}$, while the middle panel shows in more
detail the range form 10 to 100 cm$^{-1}$. Finally, the bottom panel covers the highest energy interval examined by our present 
calculations: from 100 to 200 cm$^{-1}$, which gives the largest energy considered  to be about around 300 K. We shall further discuss below the
temperature dependence in terms of the corresponding rates.

\begin{figure}
\includegraphics[scale=0.7]{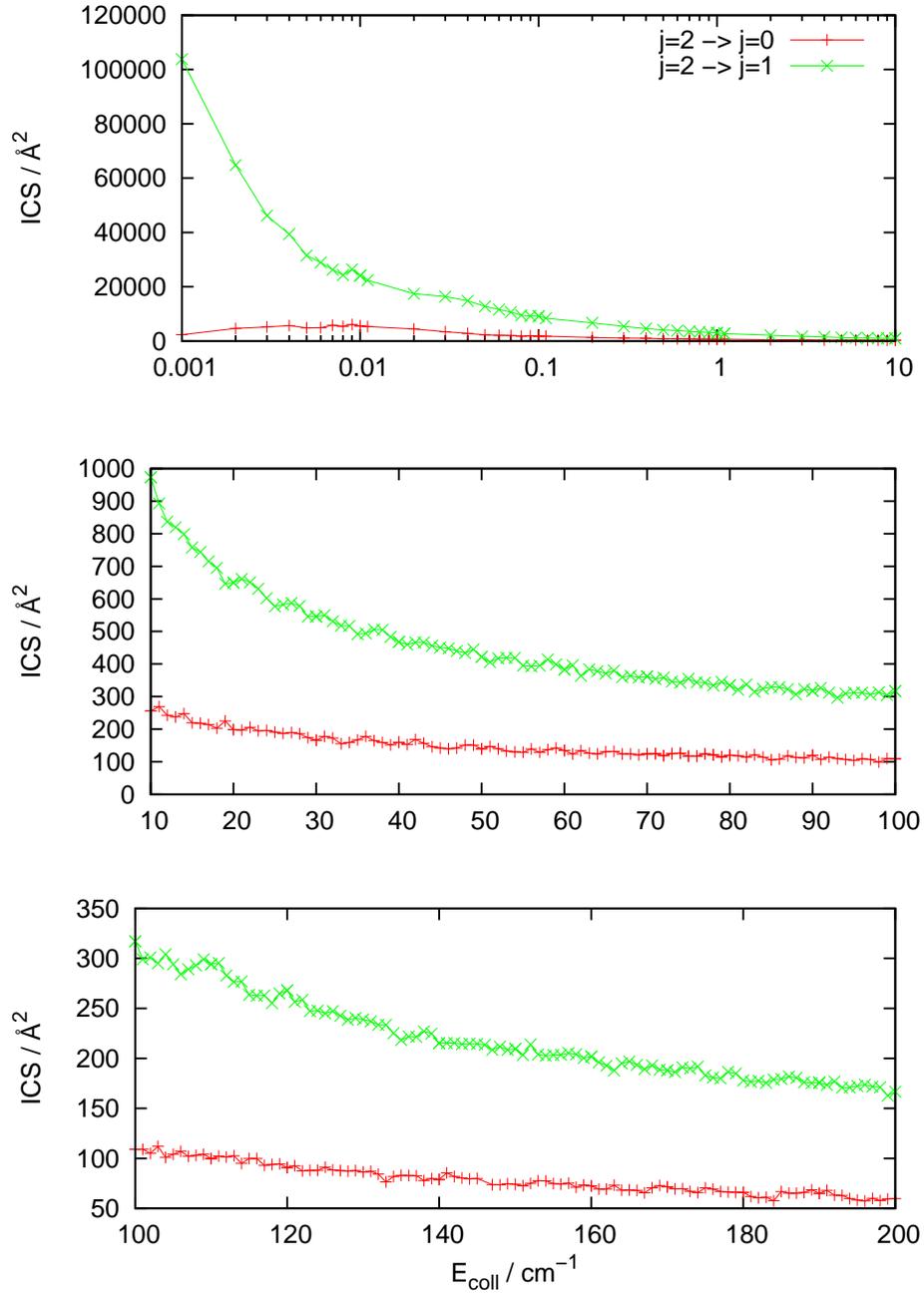}%
\caption{\label{fig:cooling2to01}Same as fig.\ref{fig:cooling1to0}, but for the rotational quenching processes from j=2 initial level.
	See main text for details.}%
\end{figure}

The following comments could be made by looking at the data, which we consider to be numerically converged with about 1\%,
over the very different energy regimes:
\begin{enumerate}
	\item at the very-low end of the energy scale we see that the collisional depletion of the j=1 level of OH$^-$ is
		indeed a very efficient process triggered by the Rb partner since the corresponding cross sections rise to more than 40$\times$10$^3$
		\AA$^2$, although they rapidly drop to still large values of $\sim$ 5,000 \AA$^2$ once the relative
		collision energy gets to be around $\sim$ 0.1 cm$^{-1}$.
	\item when the collision energy reached the values that roughly correspond to a range between 15 K and 150 K (central
		panel) we clearly see that the state-changing cross section still remains rather large, since it only 
		decreases to $\sim$ 160 \AA$^2$ by the time the translational energy gets around 150 K into the trap.
		Those values are substantially higher than those exhibited by the case  using He as a buffer gas 
		\cite{Hauser-etal:15}, where the same state-changing cross section at 100 cm$^{-1}$ turned out to be 
		around 10 \AA$^2$, i.e. about 15 times smaller than in the present case;
	\item in the highest energy range that we have considered in the present work (bottom panel of figure
		\ref{fig:cooling1to0}) we continue to see the slow decrease of the cross sections' size, together with the 
		more marked appearance of resonance structures. They are likely to be rotational Feshbach resonances due to the 
		large strength of the present coupling potential. Their detailed identification, however, is outside the scope
		(at least for now) of the present investigation and will be analysed in later studies. 
It is also worth noting that, at the highest collision energy of about 300 K, the present
		cross sections still remain as large as 60 \AA$^2$.
\end{enumerate}
The results reported by figure \ref{fig:cooling2to01} indicate in their three panels the same energy ranges which were reported
in figure \ref{fig:cooling1to0}.  They present, however, the quenching (state-changing) cross sections of OH$^-$ targets in their
j=2 initial state. Both the decays to the j=1 and j=0 states are considered.

At the lowest collision energies we find that the $\Delta$j=1 depletion cross sections are markedly larger than the ones in
figure \ref{fig:cooling1to0}: they start to be as large as 100$\times$10$^3$ \AA$^2$ at 10 cm$^{-1}$. With the same
token, these cross sections decrease slowly as the energy increases up to about 100 cm$^{-1}$, reaching there the value
of 300 \AA$^2$, which is about 1.5 larger than the state-changing cross sections, at the same collision energy, for the 
(1$\rightarrow$0) transition of \ref{fig:cooling1to0}.

It is also interesting to note that the corresponding $\Delta$j=2 cross sections for the j=2$\rightarrow$j=0 process,
also reported by figure \ref{fig:cooling2to01}, are much smaller in size throughout the same range of energies. The amount
of energy being transferred is now larger and therefore, in the expression for the corresponding cross sections
\cite{Gonzalez-Sanchez-etal:08,Tacconi-Gianturco:09} they are inversely scaled by the energy gap value, which is to be expected.
Naturally, the above scaling holds when the coupling matrix elements of the potential which connect the different states involved here are of similar strength over the whole range of coupling radial region.In general,in fact, both factors play a role . the energy gap between transition levels and potential coupling strength betheen the same.

\begin{figure}
\includegraphics[scale=0.7]{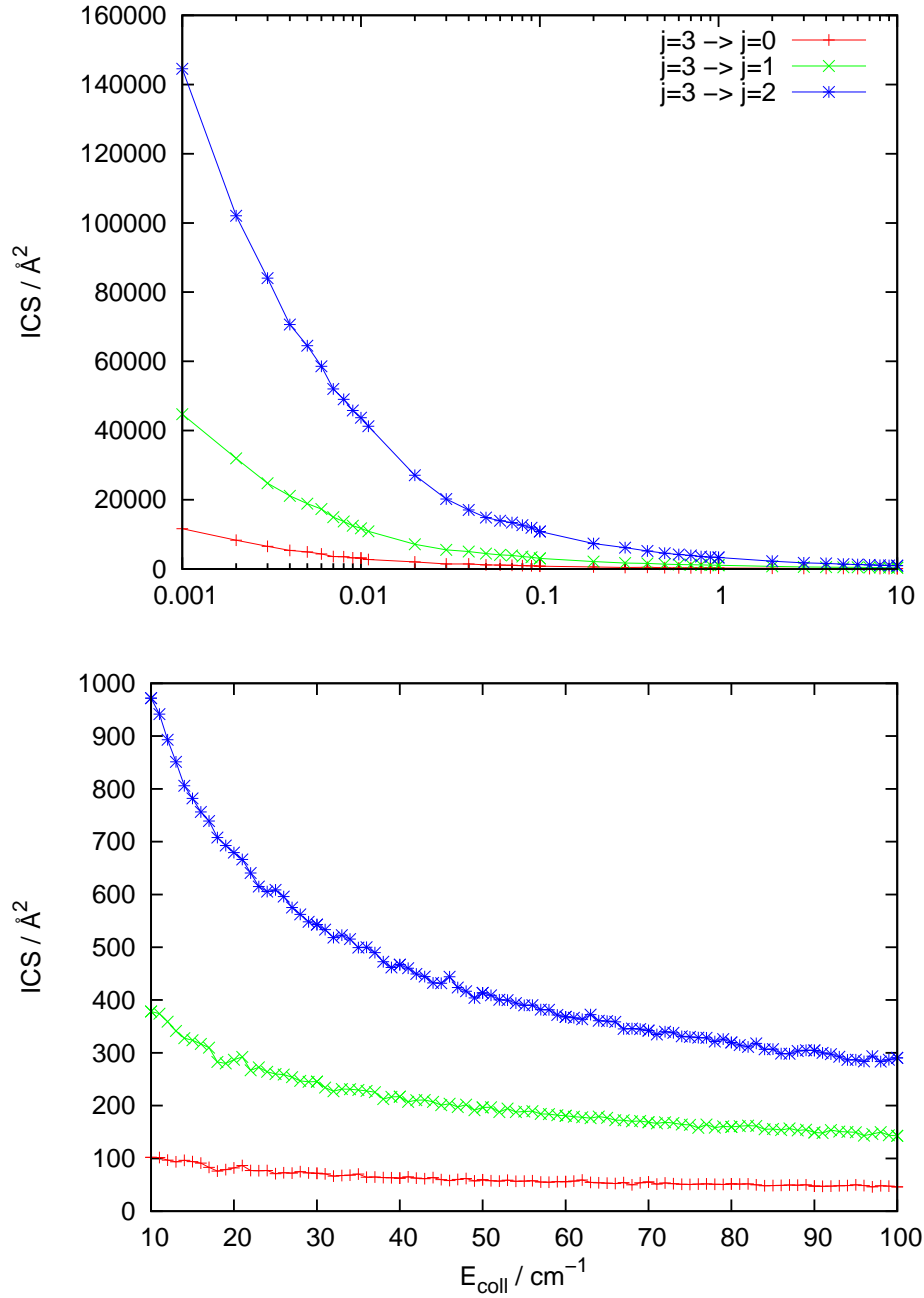}%
\caption{\label{fig:cooling3to012}Same as in figures \ref{fig:cooling1to0} and \ref{fig:cooling2to01}, but involving now the j=3 as initial 
	rotational level.
	See main text for details.}%
\end{figure}

The results given in figure \ref{fig:cooling3to012} further report , over the same range of collision energies, the behavior
of the state-changing cross sections (quenching) from the j=3 initial rotational state of the OH$^-$ molecular partner.
The size of the cross sections decreases as $\Delta$j increases from 1 , to 2 and 3, thereby producing partial
cross sections which are smaller along the series and which uniformly decrease as the collision energy increases. As an example,
at the highest considered collision energy of 140 cm$^{-1}$ the ''cooling'' cross sections vary from 220 \AA ($\Delta$j=1)
to $\sim$ 110 \AA$^2$ ($\Delta$j=2), to $\sim$ 50 \AA$^2$ ($\Delta$j=3).

It is interesting to note here that the state-changing cross sections for OH$^-$ in the j=2 state that collides with He
buffer atoms \cite{Hauser-etal:15} at a translational energy of $\sim$ 100 cm$^{-1}$ ($\sim$ 150 K) is nearly
coincident with the one from j=1 state: $\sim$ 10 \AA$^2$. The much more strongly anisotropic interaction with Rb atoms in the present study
provide at that energy much larger cross sections, which also increase from that for j=1 to that for j=2. Thus, the more
efficient role as a collisional quenchers of internal vibrational energy of Rb atoms in comparison with He atoms is clearly
confirmed by the present calculations. Besides the obvious role played by kinematic factors (i.e. by the greater mass of Rb) 
there are also the additional effects from the electronic factors which cause the overall interaction potential to be much stronger and more anisotropic
for the Rb partner atoms than for the He buffer gas atoms \cite{Gonzalez-Sanchez-Gianturco:07}. The net result, therefore, is
an overall increase of the state-changing efficiency of the laser-cooled Rb atoms with respect to the He atoms in a buffer gas.

A further comparison of the summed state-changing cross sections as one varies the collision energy is given by the data shown
by Table \ref{tab:table1}.
\begin{table}
	\begin{minipage}{\textwidth}
	\caption{\label{tab:table1}Computed state-changing (quenching) total cross sections (in \AA$^2$)
	for the first three excited rotational states of OH$^-$ as a function of the collision energy}
		\begin{tabular}{lccc}
			Energy (cm$^{-1}$) & j$_{ini}$=1  & j$_{ini}$=2  & j$_{ini}$=3 \\
			\hline
			10$^{-7}$ & 2$\times$10$^7$\footnote{data from ref. \cite{Gonzalez-Sanchez-etal:08}} & 
			4$\times$10$^7\,^{\mathrm a}$ & 10$\times$10$^7\,^{\mathrm a}$ \\
			10$^{-3}$ & 4$\times$10$^4$ & 30$\times$10$^4$ & 87$\times$10$^4$ \\
			10$^{2}$ & 1.6$\times$10$^2$ & 10.6$\times$10$^2$ & 19$\times$10$^2$ \\
		\end{tabular}
		\end{minipage}
\end{table}

The energy values reported refer to: (i) the ultracold energy regime around the nanokelvin range,
(ii) the nearly ultracold regime around the millikelvin, and, (iii) the cold regime around 150 K (bottom
line in the Table). Two features should to be clearly visible from a perusal of the data in that table:
\begin{enumerate}
	\item  as a general trend, the state-changing partial cross sections from the more excited rotational states of the target molecule into all the lower
		levels are always larger as j increases;
	\item the energy dependence for all three cases shown indicates a marked reduction of the cross sections as the
		energy increases: we increase the collision energy by nine orders of magnitude and find the cross sections being
		reduced by about five orders of magnitude;
	\item at a collision energy that roughly correspond to about 150 K (bottom line in Table) all quenching cross sections
		are still very large and increase, as the initial rotational state is increased, by more than one order  
		of magnitude.
\end{enumerate}

\begin{figure}
	\includegraphics[scale=0.7]{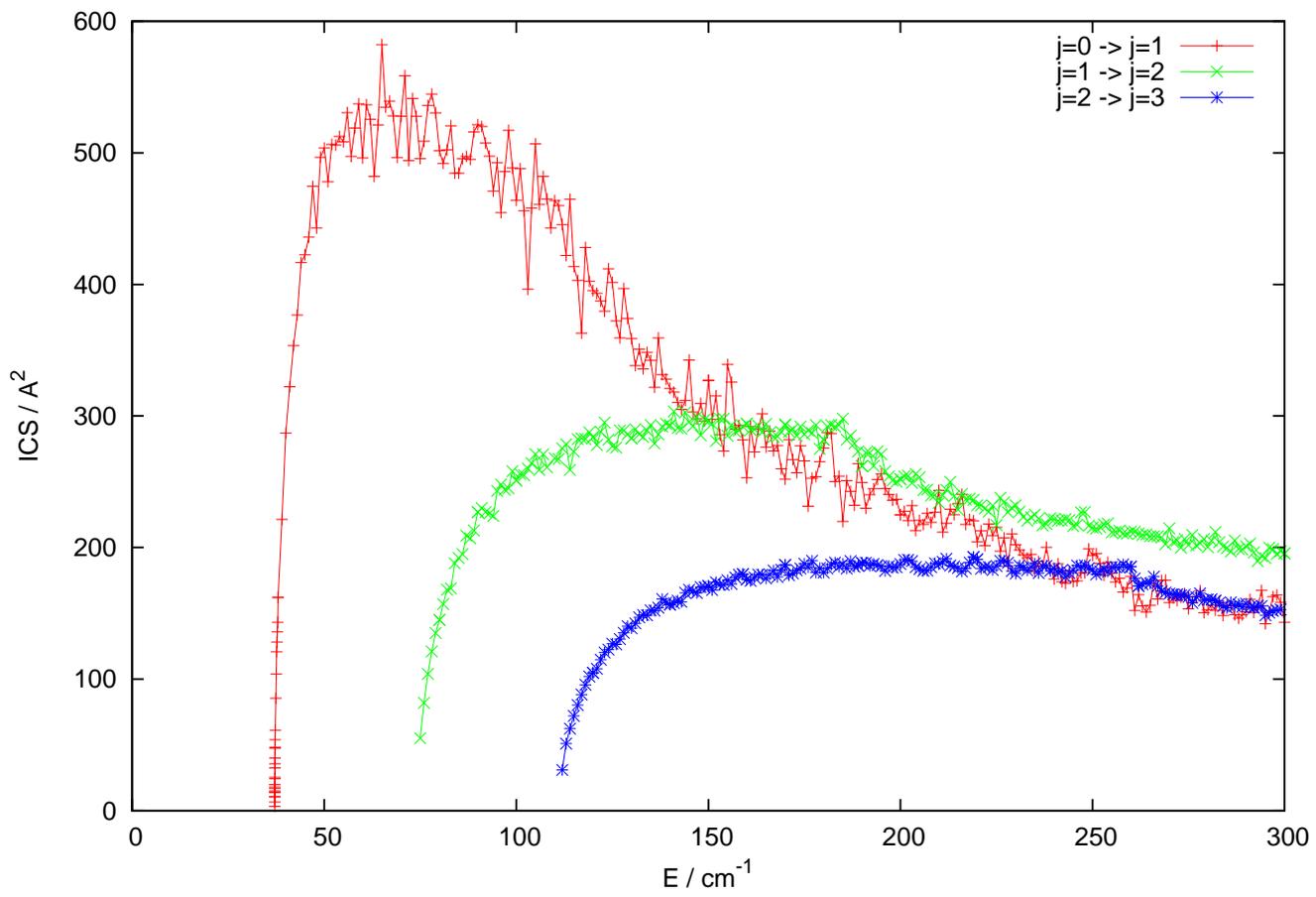}%
	\caption{\label{fig:excit}Computed state-changing (excitation) cross sections for $\Delta$j=1 processes and starting
		from the lowest three  rotational states of OH$^-$ target ion.
	See main text for further details.}%
\end{figure}

To complete the comparison between the partial cross sections from the lower rotational states of the OH$^-$ target, we further
report in figure \ref{fig:excit} the state-changing cross sections related to the excitation processes from j=0, 1 and 2 and
involving the $\Delta$j=1 transitions which turn out to be again, as we saw before for the ''cooling'' cross sections, the most important
contributions to state-changing (excitation) processes at low energies.

The data reported by figure \ref{fig:excit} are obviously covering a broader range of collision energies to allow for the successive
openings of the excitation channels.Our present findings  allow for an interesting comparison with what we have already found before with regards to
the state-changing (quenching) collisions in this system:
\begin{enumerate}
		 \item in the region between 50 and about 150 cm$^{-1}$ the molecules are taken out of the ground rotational state
			 j=0 with the largest efficiency, being then pumped up to the j=1 molecular state;
		 \item over the same energy range, however, we see that the collision probabilities for populating back the j=0
			 state from the j=1 level (mid panel of figure \ref{fig:cooling1to0}) are substantially smaller;
		 \item even if we add the additional population decay from the j=2 and 3 into the j=0 over the same range of
			 energies (mid and lowest panels in figures \ref{fig:cooling2to01} and \ref{fig:cooling3to012}) we still
			 obtain a total of state-changing probabilities which is smaller than the losses from the excitation process.
			 If we were able to further consider the actual population distributions over rotational 
states at the physical temperatures of the experiments, the
			 mismatch would become even larger: the excitation collisional losses out of the j=0 state seem
			 to remain larger than the re-populating  effects via the quenching collisions from the higher levels;
		 \item the excitation of OH$^-$ molecules out of their j=1 rotational state is also fairly large from 100 to 300
			 cm$^{-1}$ since it only decreases slowly from about 280 \AA$^2$ to about 200 \AA$^2$. However, the 
			 re-population of that state via state-changing collisions from the j=2 rotational 
level carries quite large
			 cross sections: if the j=2 state were to be substantially re-populated, it would 
contribute to the 
			 transfer of rotational population to the j=1 level with fairly large cross sections;
		 \item it is also interesting to see that, as we move to the higher collision energies the excitation cross
			 sections from j=0 to j=1 become the smallest cross sections in comparison with those which
			 excite the molecules into their higher 
			 rotational states.
	 \end{enumerate}

In conclusion, we observe here quite a complex interplay between rotational "heating" and ''cooling'' via  inelastic collisions. Only the direct 
knowledge of the state distributions for the OH$^-$ target among its accessible rotational levels could help us to better sort out the 
relative roles of the excitation/quenching cross sections we have obtained in the present work. The foregoing analysis, in fact,
simply observed the relative sizes of the partial cross sections at various energies, without as yet considering both the translational equilibrium and the internal state 
distributions. The next subsections will present the computed rates and will  partly discuss  some of the above points.

\subsection{The partial rate coefficients}

Once the individual state-changing cross sections are known, one can therefore evaluate the corresponding state-changing rates by carrying
out a convolution over the Boltzmann's distribution of relative velocities at the trap's translational temperature.

\begin{eqnarray}
	k_{j\rightarrow j'}(T) = \left( \frac{1}{\pi\mu} \right)^{1/2} \left( \frac{2}{k_B T} \right)^{3/2} \int E 
	\sigma_{j\rightarrow j'}(E) \exp{\left( -E/k_B T\right)} dE
\end{eqnarray}
Here k$_B$ in the Boltzmann's constant, E the relative collision energy and T the trap's translational temperature.

We have calculated the above rates up to about 35 K of trap temperature and employing $\sigma_{j\rightarrow j'}(E)$ values up to
350-400 cm$^{-1}$. For the integration's convergence we started with E values above threshold by about 10$^{-3}$ cm$^{-1}$ and
employed for each  k$_{j\rightarrow j'}$ rate a unevenly-spaced number of energy values which ranged up to a total of over a thousand points.Final numerical convergence was thus checked to be better than  0.01 at each T value.

\begin{figure}
	\includegraphics[scale=0.7]{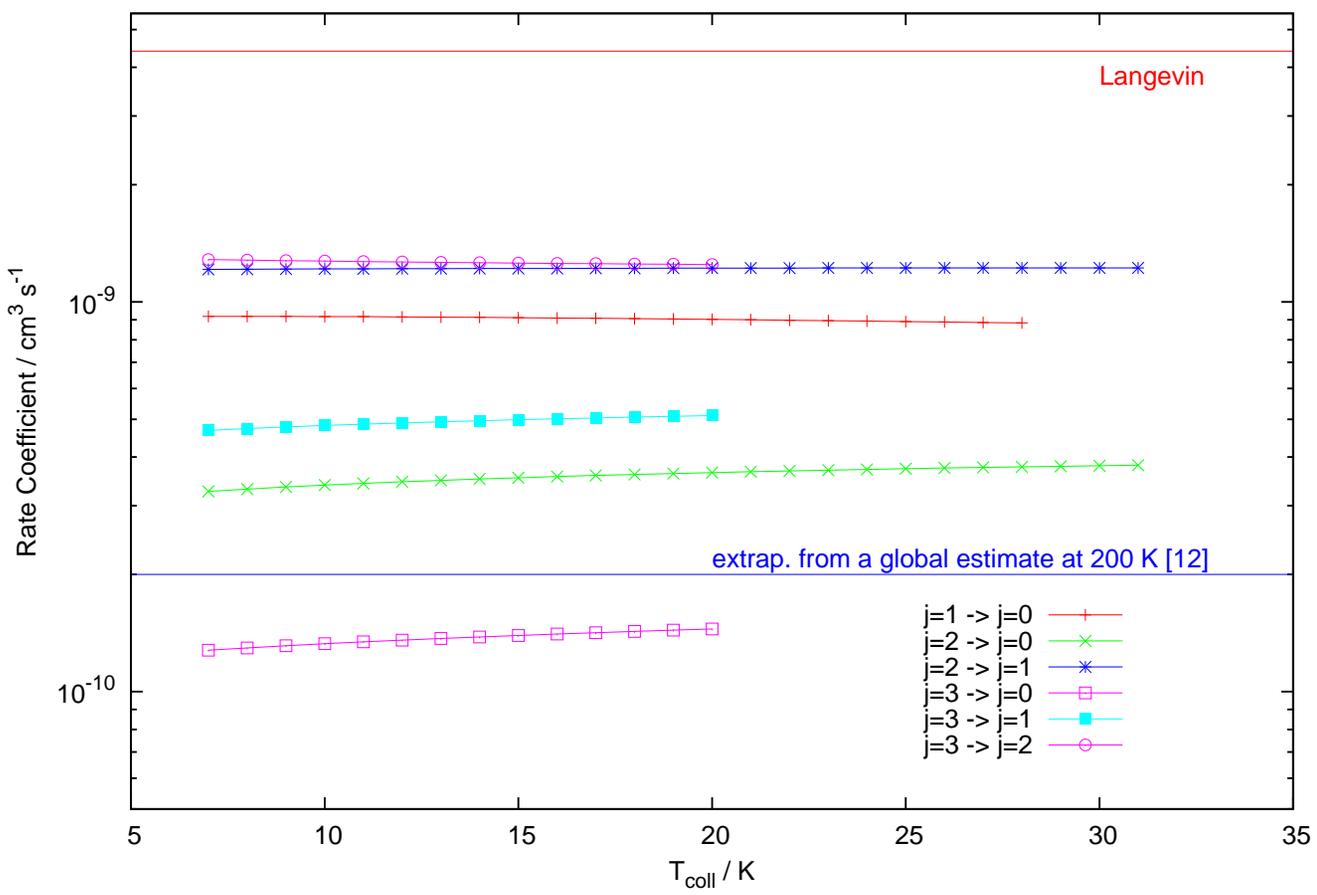}%
	\caption{\label{fig:rates}Computed state-changing rates for the rotational states of OH$^-$ in collision with rubidium atoms.
	See main text for further details.}%
\end{figure}

The results of the present calculations are given by the various curves reported in figure \ref{fig:rates}, where all the computed 
de-excitation state-changing rates are given.

The following comments could be made by examining the data in that figure:
\begin{enumerate}
	\item As already gathered from the relative sizes of the partial cross sections discussed in the previous subsection,
		we see that the largest rates (open dots and stars) correspond to state-changing (quenching) processes from
		excited rotational states of OH$^-$: j=3 and j=2, respectively. In both cases the $\Delta$j=1 process is the dominant transition.
	\item The rates for depleting the j=1 level into the j=0, given by the curve marked by vertical bars, are about 9$\times$10$^{-10}$ cm$^3$/s at 10 K and
		vary little as T increases up to 30 K. As a comparison, the same state-changing rates in the case of using He as a 
		buffer gas \cite{Hauser-etal:15} reach the value of about 5$\times$10$^{-11}$ cm$^3$/s at 10 K for the same 
		state-changing process: about one order of magnitude smaller than in the case of Rb atoms.
	\item The state-changing rates involving $\Delta$j=2 quenching transitions (the curves marked by crosses and by filled squares ), 
as well as the $\Delta$j=3 (open-squares curve) processes, yield  smaller rates, although they still appear to 
		be confined, over the examined range of temperatures, within one order of magnitude
		of the larger ones discussed above.
\end{enumerate}

It is also interesting to note that the size of the rates given by the Langevin estimates \cite{Deiglmayr-etal:12} for the present
system is T-independent and about 4.3$\times$10$^{-9}$ cm$^3$/s. One should note here,however, that such a simplified model, albeit very often used to get zeroth order estimates of 'capture' rates, 
only considers the long-range polarizability term as dominating the potential and treats all crossing trajectories as leading to reaction \cite{taylor}. It is therefore a much simpler picture than 
that developed in the present ab initio treatment and often produces a marked upper limit to the true rates.
The computed rates in figure \ref{fig:rates} are getting fairly close to that value for the state-changing rates from 
excited rotational states of OH$^-$. This feature indicates once more
the large efficiency of the state-changing rates for OH$^-$ in the case of Rb as a partner atom.

It is also interesting to note that in the earlier experiments at around 400 K ($\pm$ 200 K) \cite{Deiglmayr-etal:12}, the observed rotational population losses were attributed 
dominantly to inelastic collisions between OH$^-$ and Rb atoms and an averaged inelastic rate coefficient of about 2$\times$10$^{-10}$
cm$^3$/s was estimated. Our present findings at much lower temperatures appear, on the whole, to yield larger partial rates for 
inelastic processes, although to estimate correctly a global rate would require knowledge of the individual state distributions
in OH$^-$ at the present temperatures. Furthermore, given the achieved peak density of Rb atoms in the cloud of the experiment
of \cite{Deiglmayr-etal:12} to be 2$\times$10$^{10}$ atoms/cm$^3$, the authors estimated a collision rate of 8 s$^{-1}$.
Using the same density value for our present calculations, and assuming for the sake of the argument that the majority of the 
OH$^-$ are in their j=1 rotational state we get at 30 K a cooling collision rate of the order 18 s$^{-1}$, which is not
far from the experimental estimate  at a higher temperature, the latter being expected to decrease once the range of T values 
comes down to the aimed cooling conditions of the currently running experiments. Such a comparison, therefore, supports 
on the whole the reasonable quality of the present computational findings for the OH$^-$-Rb system with respect to what is already available from experiments. 

\begin{figure}
\includegraphics[scale=0.7]{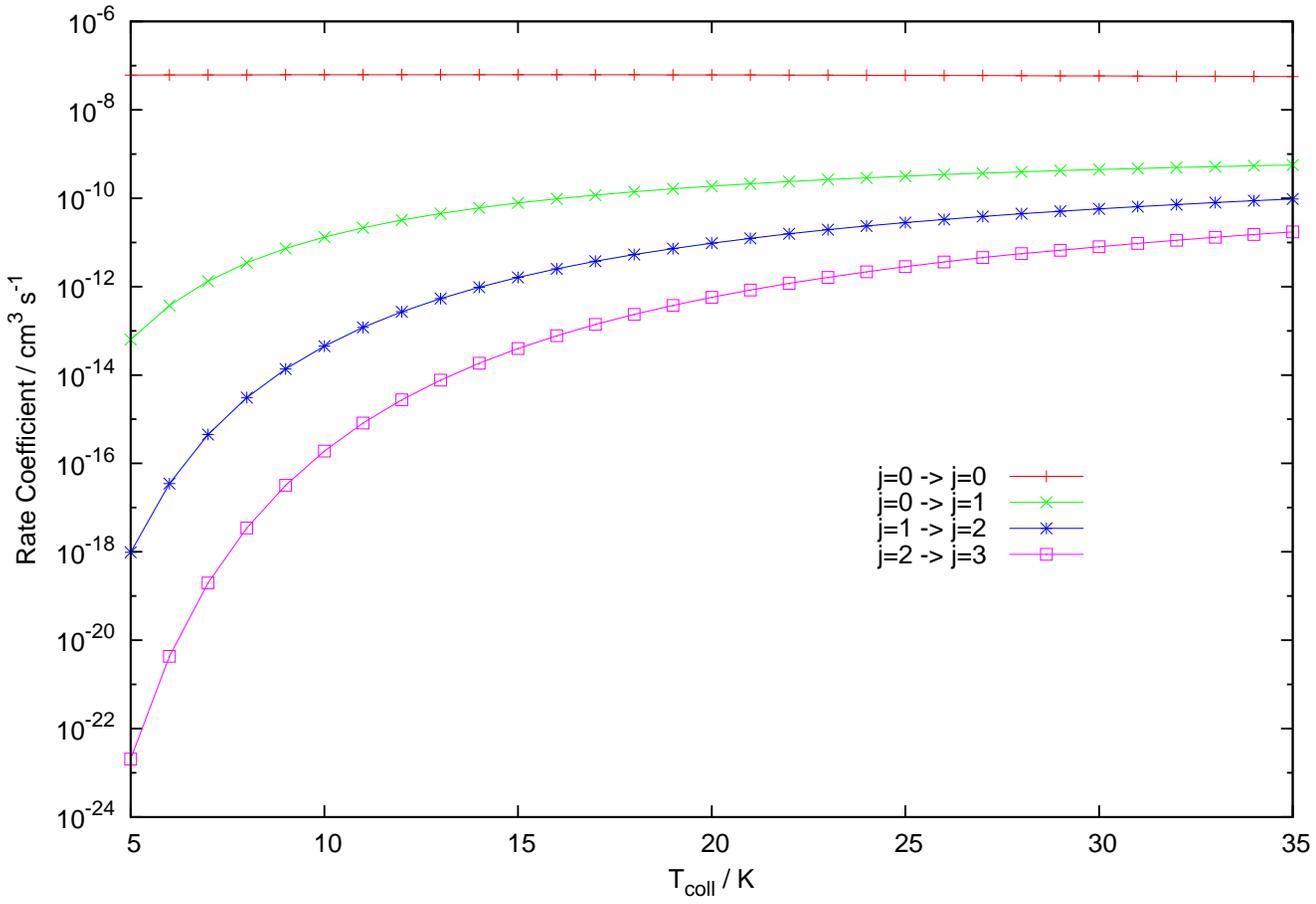}%
\caption{\label{fig:ratesexcit}%
Computed partial rates associated with rotational excitation collisional processes from the j=0, j=1 and j=2
initial molecular levels. The topmost curve corresponds to the rates associated with the elastic channel for the j=0 molecular state. 
See main text for details.}%
\end{figure}

   In order to better understand the behaviour shown by an even  broader range  of  partial collisional rates in the present system, it is interesting to further analyze the individual, partial rates associated with the  excitation cross sections discussed earlier in figure
   6. To this end, we present in figure \ref{fig:ratesexcit} the temperature dependence of the three partial rates corresponding to the excitation by one quantum of rotation for the molecular anions in their j=0 , 1 and 2 initial states. Such inelastic processes are also compared, in the same figure, with  the rates associated to the j=0 to j=0 , elastic processes.

It is interesting to note from the data in that figure that all rotational "heating" rates increase very rapidly  right above the lowest temperature considered of 5 K. Within about 10 K we see, in fact, that the smallest of the rates changes by nearly eight orders of magnitude, while the largest one changes still by four orders of magnitude. This feature clearly indicates  that to increase the relative energy between partners allows the collisions to sample  more deeply the highly anisotropic part of the PES ( e.g. see details in fig.1) and therefore each torque-changing event acquires greater importance in causing the excitation processes.
At the largest temperature sampled by our calculations, e.g. around 35 K, all the rates are now fairly large, indicating once more the efficiency in this system of the collisional energy-transfer events.
If we now look at the rate associated with the elastic channel for molecules in their ground rotational states, top curve in the figure, 
we see that the momentum transfer efficiency between the much heavier Rb atoms and the molecular partner is now much larger than in the case of the He buffer gas[13], a feature once more connected with both the changed kinematics and the stronger interaction forces.
If we were to qualitatively estimate a collision frequency value from the elastic rates by using once more the atomic density values indicated by the earlier experiments at 400 K [12], we find a qualitative value of nearly 1,000/s .This value is about 50 times larger
 than the value for the inelastic collisional quenching of the molecular anion in its j=1 rotational state.

\section{Present conclusions}

In the present study we have carried out detailed calculations for the state-changing partial cross sections involving 
excitation and deexcitation processes among the lowest four rotational levels of OH$^-$ in collision with Rb atoms. The aim has
been that of providing accurate collisional indicators for processes occurring in hybrid traps, where confined ionic
molecules are immersed in a cloud of laser-cooled heavier atoms, which are in turn employed as collisional cooling partners.

In particular, we have looked at a system for which experiments had been already done earlier \cite{Deiglmayr-etal:12}, albeit
 still at  higher temperatures than the ones which are hoped to be achieved in the current, ongoing experiments 
\cite{Wester-Weidemueller} and which we consider in the present computational study.

The evaluation of all the involved cross sections for a range of energies that go up to about 300 K, indicate the following
features in terms of their relative sizes and energy-dependence from threshold up to about to 200 cm$^{-1}$:
\begin{enumerate}
	\item both excitation and quenching cross sections are found to be, as expected, substantially larger than those
		obtained for OH$^-$ inelastic collisions with He as a buffer gas \cite{Hauser-etal:15};
	\item the quenching cross sections with $\Delta$j=1 transitions, and originating from excited rotational states
		beyond j=1, are found to be larger than the (j=1$\rightarrow$j=0) inelastic cross sections. These
		differences are more marked at the near-threshold  energies but also persist as the energy 
		is increased;
	\item the state-changing cross sections for the lowest excitation process (j=0$\rightarrow$j=1) which
		can occur for the OH$^-$ partner are not so large near threshold but rapidly become markedly larger
 as the energy 
		increases, thus suggesting a substantial effect coming from collision-induced population
 of the j=1 rotational state.
\end{enumerate}

When we further considered the corresponding state-changing partial rates, down from a few millikelvins up to about 30 K, we
also found a general behavior that was very much in  line with what we had found for the OH$^-$ + He system
\cite{Hauser-etal:15}: all the present rates also turned out to be much larger than the latter, are seen to get very close to the 
Langevin limit estimated for the present system and to exhibit a  weak temperature dependence from threshold to about 30 K, as 
indeed found earlier for the inelastic collisions with He \cite{Hauser-etal:15}.

In conclusion, the present calculations provide for the title system a quantitative assessment of the general behavior of its collisional state-changing cross sections and rates. This was done in the hope of giving a better computational support for the ongoing experiments cited before. Our findings turn out to be 
 in line with the relative trends we had   found earlier \cite{Hauser-etal:15} for the same cross sections and rates that had  involved the  He buffer gas
as a partner. The relative sizes, however, are here much larger than those for OH$^-$ with He, and confirm the  much greater
efficiency of the state-changing collisional process: a feature that, on the basis of our present discussion, is to be expected when  Rb , instead of He, is the involved partner atom. 

The extension of the present work to considering isotopic effects when using the OD$^-$+Rb system is currently under 
consideration. We also intend to extend the present analysis of rate behavior to higher temperatures, 
to allow possible closer
comparisons with the conditions under which the earlier experiments had been performed. Finally, we have also reported in figure \ref{fig:ratesexcit} 
 the size and  energy dependence of the elastic rates from the j=0 molecular state: such data are expected to give us
 a better idea on the efficiency of the sympathetic cooling of the molecular anion's translational motions under the trap conditions.

\section{Acknowledgments}
We thank Prof. Weidem{\"u}ller and his group for several clarifying discussions on the current experiments in Heidelberg. One of us
(F.C.) thanks the FWF Agency for the awarding of a postdoctoral grant during which this work was carried out. 
The computational help and support from the Computer Center of Innsbruck University is gratefully acknowledged. We are  grateful to Dr. Jesus Aldegunde for his expert advice on the evaluation of the partial rates. L.G.S. also  thanks the Spanish Ministry of Science and Innovation under grant CTQ2012-37404-C02-02. Finally, we are  grateful to the Austrian Science Fund (FWF), project P27047-N20, for the financial support of present project.

\section*{References}
%
\providecommand{\noopsort}[1]{}\providecommand{\singleletter}[1]{#1}%

\end{document}